\documentstyle[epsf,12pt,a4]{article}
\newcommand{\be}{\begin{equation}}
\newcommand{\ba}{\begin{eqnarray}}
\newcommand{\ee}{\end{equation}}
\newcommand{\ea}{\end{eqnarray}}
\newcommand{\nn}{\nonumber}

\newcommand{\tr}{\mbox{tr}}

\newcommand{\GeV}{\;\mbox{GeV}}
\newcommand{\MeV}{\;\mbox{MeV}}

\newcommand{\ol}{\overline}

\newcommand{\bbar}{\ol{B}^0(B^0)}
\newcommand{\KS}{K_{\mathrm{S}}}

\newcounter{currequation}
\newenvironment{subeqns}{\setcounter{currequation}{\value{equation}}%
\stepcounter{currequation}\setcounter{equation}{0}%
\begin{eqnarray}}{\end{eqnarray}{\setcounter{equation}{\value{currequation}}}%
}
\begin{document}
\title{
A vertex-structure model for new direct CP-violating effects in $\bbar \to \phi \KS$ and in
$B^\mp \to \pi^\mp \eta'(\eta)$
}
\author{Saul Barshay\footnote{barshay@kreyerhoff.de} and Georg Kreyerhoff\footnote{georg@kreyerhoff.de}\\
III.~Physikalisches Institut A\\
RWTH Aachen\\
D-52056 Aachen}
\maketitle
\begin{abstract}
We consider effective Lagrangian models of CP-violating vertex structure in which a
$b\to uW$ vertex, proportional to $s_{13}e^{-i\delta_{13}}$ with $s_{13}$ very small (milliweak
interaction) and $\delta_{13}$ large, is dynamically generated. A consequent, enhanced CP-violating
vertex for $b\to sg$ results in an enhanced CP-violating phase in the ratio of amplitudes for
$\bbar \to \phi \KS$. We estimate that this can significantly change the $S$ parameter from
the value expected in the standard model.
\end{abstract}
The current measurements \cite{ref1,ref2} of the $S$ parameter in the decays $\bbar \to \phi \KS$, which
at the quark level, are induced by a $b\to sg$ closed-loop (penguin) transition, with the gluon (g)
converting to an $s\ol{s}$ pair ($\phi \cong \ol{s}s$), suggest the presence of new\footnote{By new,
we mean CP-violating amplitudes in addition to those implied by the unitarized, effective CKM matrix
for quark mixing.}
CP-violating physics, which gives rise to a new phase factor in the ratio of decay amplitudes.
The tendency of $\sin 2\beta$ to deviate from the expected value of $\sim 0.7$ in the standard model
for indirect CP violation, as measured in the decays to $J/\psi \KS$ \cite{ref3,ref4}, is present in other
processes which involve the $b\to sg$ transition, in particular $\eta' \KS$ \cite{ref1,ref2}. This decay
also involves a sizable mesonic $\ol{s}s$ component amplitude when $g\to s\ol{s}$, in this case for $\eta'$
instead of $\phi$. In general, consideration of additional$^{F1}$ CP-violating physics in the $b\to sg$
transition implies the possibility of additional, related physics in the transition $b\to dg$. One
of our purposes in this paper, is to point out that there may well be an early indication of a new
physics effect in the recent data \cite{ref5,ref6} concerning direct CP-violating asymmetries in the charged
decays $B\mp\to \pi^\mp \eta'(\eta)$. This rather subtle point has not been noted before; it involves the sign of
the CP-violating asymmetry for $\pi^\mp\eta'$ relative to that for $\pi^\mp\eta$. The present data
give a suggestion that the relative sign of these two related asymmetries may be negative \cite{ref5,ref6}.
The asymmetry in partial rates $(B^- - B^+)/(B^-+B^+)$ is $0.14 \pm 0.16$ for $\pi^\mp \eta'$; it is
$-0.05 \pm 0.09$ for $\pi^\mp \eta$ (BaBar gives $-0.13\pm 0.12$) \cite{ref6}. Asymmetries with absolute
values similar to this data were predicted long ago \cite{ref7}. But the predicted asymmetries have
a positive relative sign. We show that an indication of additional physics might be contained in the possibility
that the relative sign is negative. Within a model for vertex structure, we relate hypothetical, additional CP-violating
physics in $b\to sg$ to that in $b\to dg$. The primary purpose of this paper is to describe a model of
vertex structure which gives rise to a new CP-violating 
phase $2\delta_{np}$, in the ratio of amplitudes for $\bbar\to\phi\KS$.
This new phase causes a downward 
change in the $S$ parameter, estimated below as $\sin 2\beta \to \sin(2\beta -2\delta_{np})
\cong 0.2$ \cite{ref8}. The model also gives rise to an additional
 amplitude which contributes to $B^\mp \to \pi^\mp \eta'(\eta)$.
This amplitude contains the CP-violating phase factor $e^{-i\delta_{13}}$ \cite{ref9}. In interference
with the previously calculated \cite{ref7} decay amplitude for $\pi^\mp\eta'$, the amplitude could cause
the asymmetry for $\pi^\mp\eta'$ relative to that for $\pi^\mp\eta$ to be negative. This result is a natural
consequence of the fact that the relative sign of the $\ol{s}s$ component amplitude in $\eta'$ and that
in $\eta$, is negative \cite{ref10}.\footnote{The magnitude of the $\ol{s}s$ component amplitude in
$\eta'$ is about $1.4$ times larger than that in $\eta$, for an $SU(3)$ singlet-octet mixing angle
of $\sim -20^\circ$ \cite{ref10}, and is about twice as large as the $\ol{u}u$ ($\ol{d}d$) component
in $\eta'$.}

The empirical pattern of mixing in the quark sector \cite{ref9} indicates that there is a single,
intrinsically very small mixing angle $s_{13} < 10^{-1}$ of either of the other two mixing angles $s_{23},s_{12}$.
This angle is associated with the KM CP-violating \cite{ref11} phase $\delta_{13}$,\footnote{The standard
form of the CKM matrix \cite{ref9}, which has the $b\to uW$ vertex element parameterized as
$s_{13}e^{-\delta_{13}}$, can be presented in alternative forms \cite{ref11}. The point of view of this
paper is that the standard form incorporates the possibility of a dynamical generation of the $b\to u$
element. The appearance of the combination $s_{13}e^{i\delta_{13}}$ in small parts of other elements
is a consequence of enforcing unitarity of the effective CKM matrix.}
which as presently approximately determined, is a large phase ($>45^\circ$) \cite{ref9}. This coincidence
of a large phase and a very small mixing angle could suggest that there exists some kind of milliweak
interaction which grossly violates CP invariance \cite{ref12}. This interaction could dynamically give rise
to a sizable CP-violating phase in $b\to uW$, which appears together$^{F3}$ with a very small mixing
parameter associated with the milliweak strength. In order to obtain the phenomenological, numerical
results below, which are directly relevant to the above anomalous properties of CP violation in B decays,
we assume that if there is an underlying dynamical structure in the $b\to uW$ transition, then there
will be evidence for this from additional vertices which violate CP invariance explicitly. The model takes
as a specific, simple example, the $b\to uWg$ vertex shown in Fig.~(1a). Clearly then, the diagrams in Fig.~(1b)
induce new, related vertices for $b\to sg$ and for $b\to dg$. We estimate these now, and apply the results
to $\bbar\to\phi \KS$, and then to $B^\mp \to \pi^\mp \eta'(\eta)$. 

Our first example of a CP-violating, effective interaction for the vertex in Fig.~(1a) is written as
\be
H_{\mathrm{eff}} = e s_{13} e^{-\delta_{13}} \left(\frac{g_s g}{M^3}\right) \ol{\psi}_u
\lambda^a \psi_b G^a_{\mu\nu} W^{\mu\nu} + {\mathrm{herm.~conj.}}
\ee
where $g^2/8M_W^2 = G_F/\sqrt{2}$, the Fermi coupling, $g_s^2/4\pi = \alpha_s$, the QCD coupling, $M$ is a mass
parameter, $e = \pm 1$, the $\psi$ are the $b$ and $u$ quark fields, and $G_{\mu\nu} = (\partial_\nu G_\mu - 
\partial_\mu G_\nu)$ and $W^{\mu\nu}$ are the gluon and $W$ field
tensors, respectively (we do not consider the quadratic terms). The $\lambda^a$ are the $SU(3)$-color matrices \cite{ref13}, $a=1\ldots 8$. Using
$M=m_W$ in (1), the new effective vertex for $b\to sg$, estimated from the graph in Fig.~(1b) in the
approximation of retaining the leading terms in the intermediate-state $W$ four-momentum, and of taking
the integration variable from zero to an upper limit of $m_W^2$ is \footnote{The $S$-matrix is calculated
as $-i(2\pi)^4 \delta^4(P_{\mathrm{in}} - P_{\mathrm{fin}})V$, where $P_{\mathrm{in}}$ and
$P_{\mathrm{fin}}$ denote initial and final total energy-momentum. Power counting gives a quadratic
integral over four-momentum. The cut-off is assumed to be a consequence of form factors in $H_{\mathrm{eff}}$.}
\be
V_{\mathrm{new}} \cong e s_{12} s_{13} e^{-i\delta_{13}}\sqrt{2}G_F \left(\frac{g_s}{16\pi^2}\right)(0.43 m_W)
\ol{s}\lambda^a (q\!\!\!/ \epsilon\!\!\!/ - \epsilon\!\!\!/ q\!\!\!/)b_R
\ee
where $b_{R,L} = 1/2 (1\pm \gamma_5)b$ denote right and left-handed components formed from $b$ spinors,
and $q_\mu$ and $\epsilon_\mu$ are the gluon four-momentum and polarization vector, respectively ($q\!\!\!/
= \gamma^\mu q_\mu$). This vertex is enhanced by a numerically large ratio of masses, relative to the
dominant contribution to the standard-model, penguin-type vertex $b\to sg$ from the $c$ and $u$ quark intermediate
states in particular, $(V(c)+V(u))$, with \cite{ref14,ref15}
\begin{subeqns}
V(c)=s_{23} \sqrt{2} G_F \left(\frac{g_s}{16\pi^2}\right) F_1(m_W^2/m_c^2) \ol{s}\lambda^a(q^2\epsilon\!\!\!/ -
q\cdot\epsilon q\!\!\!/)b_L\\
V(u)=s_{12}s_{13} e^{-i\delta_{13}}
	\sqrt{2} G_F \left(\frac{g_s}{16\pi^2}\right) F_1(m_W^2/m_c^2) \ol{s}\lambda^a(q^2\epsilon\!\!\!/ -
q\cdot\epsilon q\!\!\!/)b_L
\end{subeqns}
and
\ba
F_1(m_W^2/m_c^2) \cong \frac{2}{3} \ln \left(\frac{m_W^2}{m_c^2}\right) \cong 5.5\nn\\
F_1(m_W^2/m_u^2) \cong \frac{2}{3} \ln \left(\frac{m_W^2}{m_u^2}\right) \cong 13.8\nn
\ea
for $m_W\cong 80\GeV$, $m_c\sim 1.25\GeV$, $m_u\sim 2.5\MeV$.
Using the empirical \cite{ref9} mixing angles with $s_{12}s_{13} \sim (s_{12}^2/2.5)s_{23}\sim 0.02 s_{23}$,
$(c_{12},c_{23},c_{13}\sim 1)$,
and using $\delta_{13} \sim 90^\circ$, the magnitude of the CP-violating phase of the full vertex, $\delta$,
which comes from the term (3b), is at most of the order of $(0.02) (13.8/5.5)\sim 0.05$, quite small.
This is the origin of the result that the S parameter for $\bbar\to \phi\KS$ should not deviate by more
than about $0.07$ from $\sin 2\beta = 0.7$, where $S_{\phi\KS} = -{\mathrm{Im}} e^{-i(2\beta + 2\delta)}$. Apart from
the common factor of $s_{12}s_{13} e^{-i\delta_{13}}$, the effective magnitude of the new vertex is
enhanced by a factor of order $(0.43m_W/\sqrt{q^2})/F_1(m_W^2/m_u^2) \cong 86/13.8 \cong 6$, as is evident by
comparision of the dimensional quantities in Eq.~(2) and Eq.~(3b), for relevant $q^2>0$ \cite{ref15}.
This gives rise to a new, sizable CP-violating phase $\delta_{np}$ in the ratio of amplitudes
$A(\ol{B}^0\to \phi \KS)/A(B^0\to \phi\KS) = e^{2i\delta_{np}}$, where for $e=-1, \sin\delta_{13}\sim 1$,
and a typical $q^2 \sim (0.4 \GeV)^2$, we illustrate a numerical value
\ba
\delta_{np}&\sim& \tan^{-1}\left\{ \frac{s_{12}s_{13}(0.43m_W/\sqrt{q^2})}{s_{23}F_1(m_W^2/m_c^2)}\right\} \sim
\tan^{-1}\left\{ \frac{86}{5.5} \times \frac{s_{12}^2}{2.5}\right\} \nn\\
&\cong&\tan^{-1}(6 s_{12}^2) \cong
\tan^{-1}(0.3) \cong 17^\circ
\ea
For this example, the S parameter becomes
\be
S_{\phi\KS} = \sin(2\beta - 2\delta_{np}) \sim \sin 11^\circ \sim 0.2
\ee
An empirical average of $0.34 \pm 0.20$ \cite{ref1} is recently updated as about (preliminary) 
 $0.47 \pm 0.19$ \cite{ref2}. 
Also deviating downward at BaBar, $S_{\eta'\KS}=0.36\pm 0.13$ \cite{ref2}. If these sizeable deviations
are established by further data, there must be additional CP-violating physics.
The example of a downward change in the $S$ parameter in Eq.~(5) is the main result of this model.
Below, we show in addition that the related
new vertex for $b\to dg$, with sign and enhanced magnitude now fixed, can contribute a dominant negative
amplitude in $B^\mp\to \pi^\mp \eta'$, which can result in a CP-violating asymmetry opposite in sign to
that in $B^\mp\to \pi^\mp\eta$. Note that the vertex in Fig.~(1a) gives rise to an amplitude for $b\to uW$,
via the graph in Fig.~(2). This effective vertex, again estimated in the approximation$^{F4}$ 
of retaining the leading term in the intermediate-state gluon four-momentum, and of taking the
integration variable from zero to $m_W^2$ is
\be
V\cong es_{13}e^{-i\delta_{13}}\left(\frac{g}{2\sqrt{2}}\right)\left(\frac{8\sqrt{2}}{3}\right)
\left(\frac{g_s^2}{16\pi^2}\right) \ol{u}\left\{\frac{\epsilon_W\!\!\!\!\!\!\!\!/\;\;\;\; p_W\!\!\!\!\!\!\!\!/\;\;\;\; - 
p_W\!\!\!\!\!\!\!\!/\;\;\;\;\epsilon_W\!\!\!\!\!\!\!\!/\;\;\;\;}{m_W}
\right\}b
\ee
This vertx has a small strength magnitude relative to the $s_{13}e^{-i\delta_{13}}(g/2\sqrt{2})$ of the standard
CKM element, of the order of $\alpha_s(8\sqrt{2}/12\pi)\sim 0.06$, for $\alpha_s\cong 0.2$.

We give an additional example of an effective interaction for the vertex in Fig.~(1a). This interaction 
allows us to 
explicitly illustrate the dynamical generation of a vectorial term proportional to $s_{13}e^{-i\delta_{13}}$,
which contributes to the $b\to uW$ element of the effective CKM matrix. This CP-violating effective
interaction\footnote{We note
that a hypothetical interaction of this form, with the changes $u\to b$, $W\to Z$, $M=m_Z$ and a
real coupling, has been studied in connection with motivating searches for CP-violating effects
in 3 and 4-jet decays of the $Z$. See reference 16 and earlier references therein.}
is 
\be
H'_{\mathrm{eff}} = e' s_{13}e^{-i\delta_{13}}\left(\frac{g_sg}{M^2}\right)i \ol{u}_L \lambda_a \gamma_\mu b_L
G_a^{\mu\nu}W_\nu + {\mathrm{herm. conj.}}
\ee
The effective vertex analogous to (6) is, for $M=m_W$,
\be
V' \cong e' s_{13} e^{-i\delta_{13}}\left(\frac{g}{2\sqrt{2}}\right) (8\sqrt{2})\frac{g_s^2}{16\pi^2}\ol{u}_L
\epsilon_W\!\!\!\!\!\!\!\!/\;\;\;\; b_L
\ee
The dynamically-generated strength magnitude is then about $18\%$ of the standard CKM element ($e'=1$).
The vertices for $b\to s(d)g$ generated from (7) via graphs as in Fig.~(1b), are small; they do not
contain the enhancement factor proportional to 
$(m_W/\sqrt{q^2})$.\footnote{Power counting gives a logarithmic integral over four-momentum.} 

To clearly present the issue of the CP-violating asymmetries in the partial rates for $B^{\mp}\to \pi^{\mp}\eta'(\eta)$,
we briefly summarize the basis \cite{ref7} for these decays being likely the first place, together
with $B^{\mp}\to K^{\mp}\eta$, \cite{ref17,ref5,ref6} where direct CP violation is observed in the decays
of a charged particle. (We also update a model parameter.) The phenomenological ``bare''\footnote{The ``bare''
amplitudes  become physical amplitudes after acquiring calculated \cite{ref7} phases from strong final-state
interactions involving inelastic transitions $\pi^\mp \eta_c \leftrightarrow \pi^\mp \eta(\eta')$.}
amplitudes, corresponding to the quark diagrams in Fig.~(3a,b) are
\ba
A_{\eta} &=& (0.57)(0.25) (f_\pi a_1 + f_\eta a_2) \left(m_B^2G_F/\sqrt{2}\right)(s_{13}e^{-i\delta_{13}})\nn\\
&\sim& (0.14)(a_1+a_2)f_\pi \left(m_B^2G_F/\sqrt{2}\right) (s_{13} e^{-i\delta_{13}})\nn\\
A_{\eta'} &=& (0.4)(0.25) (f_\pi a_1 + f_{\eta'} a_2) \left(m_B^2G_F/\sqrt{2}\right)(s_{13}e^{-i\delta_{13}})\nn\\
&\sim& (0.1)(a_1+a_2)f_\pi \left(m_B^2G_F/\sqrt{2}\right) (s_{13} e^{-i\delta_{13}})
\ea
The first factor is the weight of the $\ol{u}u$ component in $\eta$, $\eta'$, for a mixing angle of $-20^\circ$ $^{F2}$
\cite{ref10}. The following two factors occur in the model \cite{ref18,ref19} for exclusive $B$ decays which we 
have used \cite{ref7} in going from quark diagrams to the matrix elements for physical hadrons. The parameters
$a_1\sim 1$ and $a_2\sim 0.2 (\sim \alpha_s)$, refer to the strength of two parts of the model's effective
weak interaction involving four quarks \cite{ref18}, where the terms proportional to $a_2$ are a simple
phenomenological parametrization of effects which are assumed to arise from hard-gluon corrections to 
the basic charged-current interaction (terms in $a_1$).\footnote{The original model \cite{ref18} used 
$a_2\sim -0.2$ for exclusive decays of heavy quarks. Subsequent arguments \cite{ref19}, suggest that
the parameter is $\sim +0.2$ at a momentum scale relevant for the $b$ quark. In our early calculation \cite{ref7}
for $\pi^\mp \eta(\eta')$, we used $a_2\sim -0.2$. Here, we use $a_2\sim 0.2$ and give resulting estimates
for rates and asymmetries.} The $f$ are pseudoscalar-meson decay
constants, which arise in going to physical hadrons, as explicitly indicated in Figs.~(3a,b,c). 
In our numerical estimates we use $f_\eta=f_{\eta'}=
f_\pi$ \cite{ref20}. The decay model \cite{ref18} involves ``overlap'' parameters $F^{B\to \pi,\eta,\eta'}$,
which arise as also indicated in Figs.~(3a,b,c). For our present numerical estimates, it suffices to approximate
these three measures of matrix-element strength by a single value, $\sim 0.25$ \cite{ref7}. We recall
that the dynamical origin of the CP-violating asymmetries has two essential elements. There is the interference
of the amplitudes in Eqs.~(9) (containing the KM CP-violating phase $\delta_{13}$), with an amplitude $A_{\eta_c}
\propto a_2 s_{23}$, corresponding to the $\ol{c}c$ diagram in Fig.~(3b). Then there is the fact that the physical
decay amplitudes contain calculated \cite{ref7} phases from strong final-state interactions involving inelastic
transitions $\pi^\mp \eta_c \leftrightarrow \pi^\mp \eta (\eta')$. These transitions depend upon coupling parameters
$g_{\eta_c\eta}$ and $g_{\eta_c \eta'}$, whose (similar) absolute magnitude is determined \cite{ref7} from
data on $\eta_c \to \eta(\eta') \pi^+\pi^-$.\footnote{With $a_2>0$, $g_{\eta_c\eta} \sim g_{\eta_c\eta'}<0$
gives negative asymmetries for $\pi^\mp\eta$ and for $\pi^\mp\eta'$, (since in this model, the asymmetry
sign depends upon the sign of the product $a_2 g_{\eta_c\eta(\eta')}$). This sign for $g_{\eta_c\eta(\eta')}$
is the same as that for $g_{\eta\eta'}$, as given by an $SU(3)$ argument \cite{ref7}. It is possible to obtain
a positive asymmetry for $\pi^\mp\eta'$ by having $g_{\eta_c\eta'}>0$. This paper illustrates a different,
dynamical possibility for having asymmetries of opposite sign in $\pi^\mp\eta'$ and $\pi^\mp\eta$, within
the framework of the same new CP-violating physics as discussed for $\phi \KS$.} 
Our results for branching ratios ${\mathrm{(b.r.)}}_{\eta(\eta')}$ and
asymmetries $a_{\eta(\eta')}$ (updated $^{F8,F9}$ from reference 7) are
\ba
{\mathrm{(b.r.)}}_{\eta} \sim 4.8 \times 10^{-6} &\;\;\;\;\;\;\;\;\;\;& a_\eta \sim -(3-7)\% \nn\\
{\mathrm{(b.r.)}}_{\eta'} \sim 2.3 \times 10^{-6} &\;\;\;\;\;\;\;\;\;\;& a_{\eta'} \sim -(5-12)\% 
\ea
The larger asymmetry magnitudes are for $f_{\eta_c} \sim 310\MeV$, instead of $f_{\eta_c}\sim f_\pi$ \cite{ref7}.
The branching ratios agree with the data \cite{ref21}.

These results could be changed by the presence of a significant amplitude arising from the quark diagrams in Fig.~(3c), for example.\footnote{
In our numerical estimates, we take the same amplitude for $g\to s\ol{s},u\ol{u},d\ol{d}$. We do
not consider effects of $g\to c\ol{c}$, assumed to be relatively suppressed, upon $A_{\eta_c}$ \cite{ref7}.} This amplitude is enhanced relative to a standard-model magnitude, by the same new physics as we
have discussed above for the possible enhanced, amplitude phase in $\phi\KS$. We use the
enhanced transition for $b\to dg$ that is implied by the enhanced transition for $b\to sg$ (Fig.~(1b)) . 
The enhanced amplitude
from Fig.~(3c) contributes mostly$^{F2}$ to $\pi^\mp \eta'$ via the positive $\ol{s}s$, $\ol{u}u$ and
$\ol{d}d$ component amplitudes in $\eta'$; this amplitude is opposite in sign $(e=-1)$ to the $A_{\eta'}$
in Eq.~(9). Added to $A_{\eta'}$, it can change the sign of a new full amplitude $A'_{\eta'}$. Then, $A'_{\eta'}$
in interference with $A_{\eta_c}$, gives a changed sign for the asymmetry in $\pi^\mp \eta'$. We calculate a
rough numerical example to illustrate that the enhanced magnitude of the amplitude from Fig.~(3c) could
effect this change of sign.
Leaving out factors in the physical amplitude derived from Fig.~(3c) which are in common with those of $A_{\eta'}$
in Eq.~(9) (from Figs.~(3a,3b)), then simply changing the sign of $A_{\eta'}$, while retaining the magnitude
requires relevant factors from the enhanced amplitude to satisfy\footnote{In the diagrams in Figs.~(3c) there
is a $(-1)$ from the propagator of the gluon with $q^2>0$, and a compensating $(-1)$ from the subsequent
gluon-exchange interaction (with $\tilde{q}^2<0$) between quark and antiquark, which is necessary to neutralize
color. The virtual (real amplitude) intermediate state is estimated. There are additional diagrams;
the present estimate assumes no overall suppression (apart from the further factor of $\alpha_s$).}
\ba
&&\sqrt{2}\left(\frac{g_s}{2}\right)\left(\frac{g_s}{16\pi^2}\right)\left(\frac{g_s}{2}\right)^2\left(\frac{1}{\pi^2}\right)
\left(\frac{16}{3}\right)\left(\frac{0.43 m_W}{\sqrt{q^2}}\right)\left(\frac{4}{\sqrt{6}}\right) \nn\\
&& \cong 2\times \left( \frac{1}{\sqrt{2}}(a_1+a_2)\left(\frac{1}{\sqrt{6}}\right)\right)
\ea
For $\alpha_s \cong 0.2$, $\sqrt{q^2} \sim 0.4 \GeV$, $(a_1+a_2)\sim 1.2$, Eq.~(11) reads numerically
\be
\frac{1}{\sqrt{2}}\left(\frac{4(0.2)^2}{3\pi^2}\right)(86)\left(\frac{4}{\sqrt{6}}\right) \sim 2\times\left(
\frac{1}{\sqrt{2}}(1.2)\left(\frac{1}{\sqrt{6}}\right)\right)
\ee
or
$$
0.53\sim 0.69
$$
The proximity of the two sides of Eq.~(12), which arise from different dynamics, is a consequence
of the enhancement factor $(0.43m_W/\sqrt{q^2})$, and also of the large contribution from the
$\ol{s}s$ amplitude in $\eta'$, $(2/\sqrt{6})$, plus the contributions from $\ol{u}u$ and $\ol{d}d$,
$(1/\sqrt{6})$ each$^{F2}$. The amplitude from Fig.~(3c) is enhanced despite the explicit inclusion
in our estimate of a factor $(g_s/2)^2 (16/3\pi^2) = (16/3\pi) \alpha_s \sim 0.34$, which
occurs in order to neutralize the color of the $q\ol{q}$ pair created by the gluon from the
$b\to dg$ transition. The factor $(16/3) = \tr(\tilde{\lambda}_a\lambda_a)^2$ is the
relevant overall color factor, where $\tilde{\lambda}_a$ is the transposed matrix. The sign
of $A_{\eta'}$ is changed. Therefore the sign of the asymmetry $a_{\eta'}$ in Eq.~(10) is
changed to positive. The new amplitude decreases $A_\eta$ somewhat. (The $\ol{u}u$,
$\ol{d}d$ and $\ol{s}s$ component amplitudes in $\eta$ are of the same magnitude
$\sim 1.4/\sqrt{6}$, but the $\ol{s}s$ component is negative.\footnote{Differing interference
effects in exclusive $B$ decays, due to the different sign of the $\ol{s}s$ component
in $\eta$ and $\eta'$, were discussed in reference 22.}) The branching
ratio is reduced by a factor of about $0.3$, but the sign of the asymmetry does not change.
Our purpose here is to supplement the main result of the paper concerning the change of the
$S$ parameter for $\phi \KS$ due to an enhanced CP-violating vertex for $b\to sg$, with
an indication of a possible effect from the related enhanced vertex for $b\to dg$. If
the contribution of the new amplitude from Fig.~(3c) is smaller than suggested in Eq.~(12),
there is another possibility for explaining a positive sign of $a_{\eta'}$ $^{F9}$ should
experiment confirm this \cite{ref5,ref6}, and the branching ratios in Eq.~(10) are little
affected.

In summary, following the idea that a $b\to uW$ vertex, proportional to $s_{13} e^{-i\delta_{13}}$,
is dynamically generated, we have given examples of structure which give rise to related,
enhanced CP-violating vertices for $b\to sg$ and $b\to dg$. Applying the model vertex to
$\ol{B}^0(B^0)\to \phi \KS$, an estimate indicates that an enhanced CP-violating phase is
induced in the ratio of these decay amplitudes. Thus, the $S$ parameter can be significantly
changed from the standard-model expectation.

\clearpage
\section*{Figures}
\begin{figure}[h]
\begin{center}
\mbox{\epsfxsize 13cm \epsffile{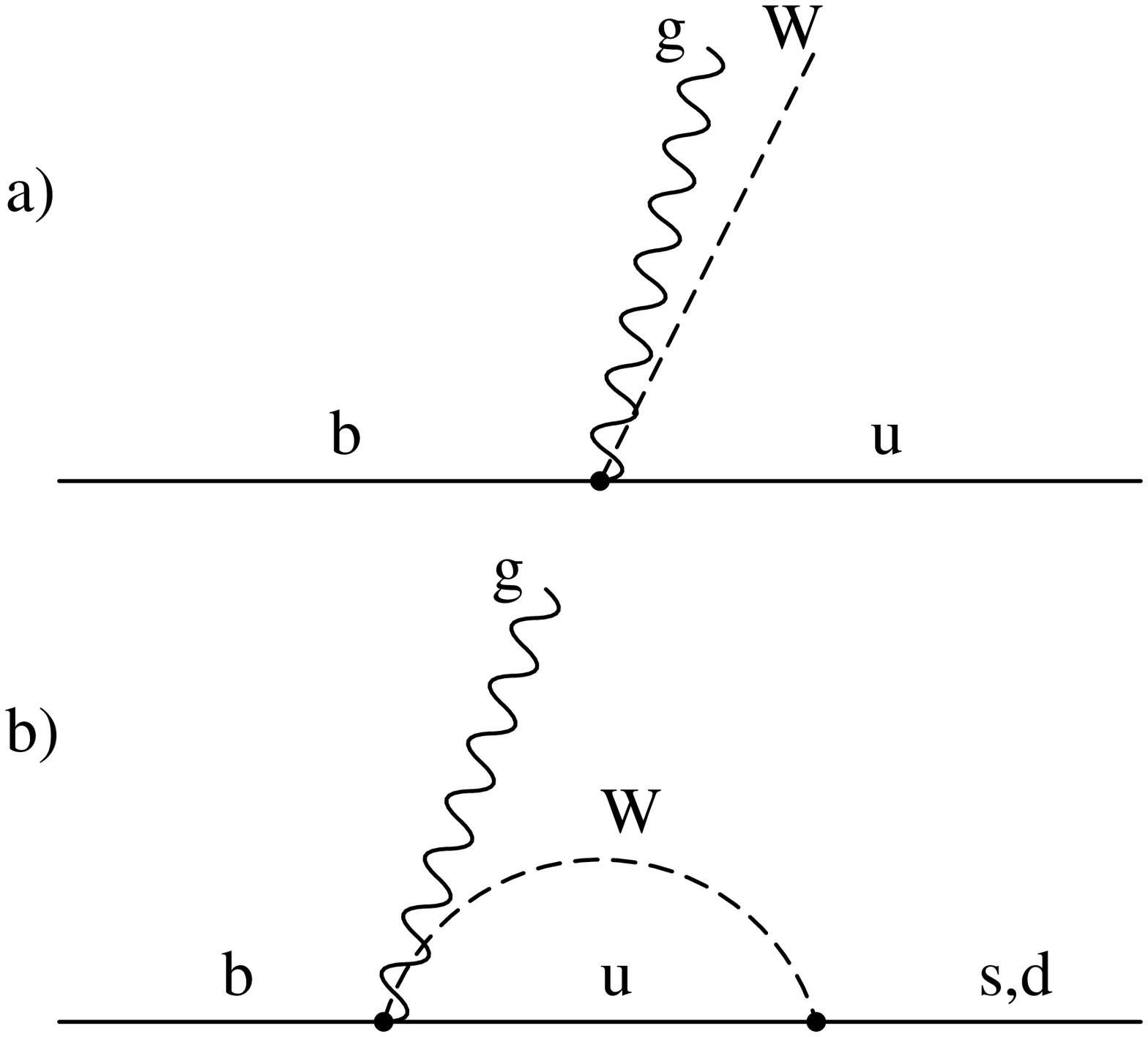}}
\caption{\small a) The vertex in Eq.~(2) that is the basis for the structure discussed in this paper. \newline b) 
Effective vertices for $b\to s(d)g$ generated from the vertex in a) and a weak interaction vertex.}
\end{center}
\end{figure}
\clearpage
\begin{figure}[h]
\begin{center}
\mbox{\epsfxsize 13cm \epsffile{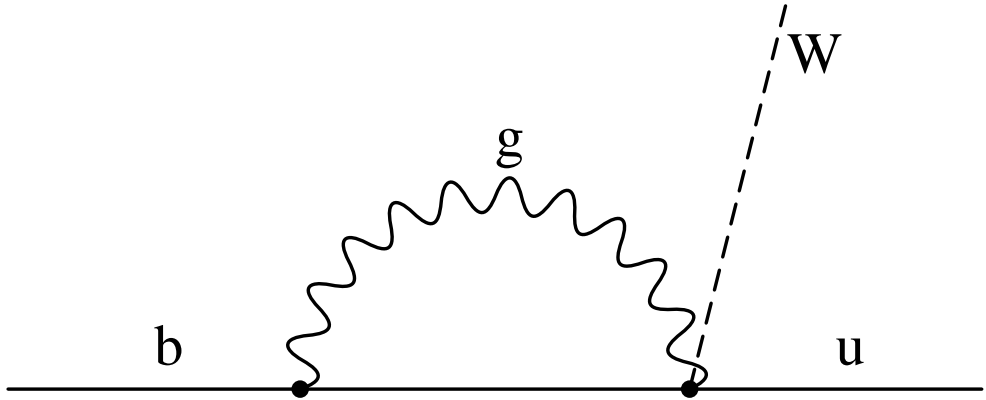}}
\caption{\small An effective vertex for $b\to uW$ generated from the vertex in Fig.~(1a), proportional
to $s_{13}e^{-i\delta_{13}}$.}
\end{center}
\end{figure}
\clearpage
\begin{figure}[h]
\begin{center}
\mbox{\epsfxsize 13cm \epsffile{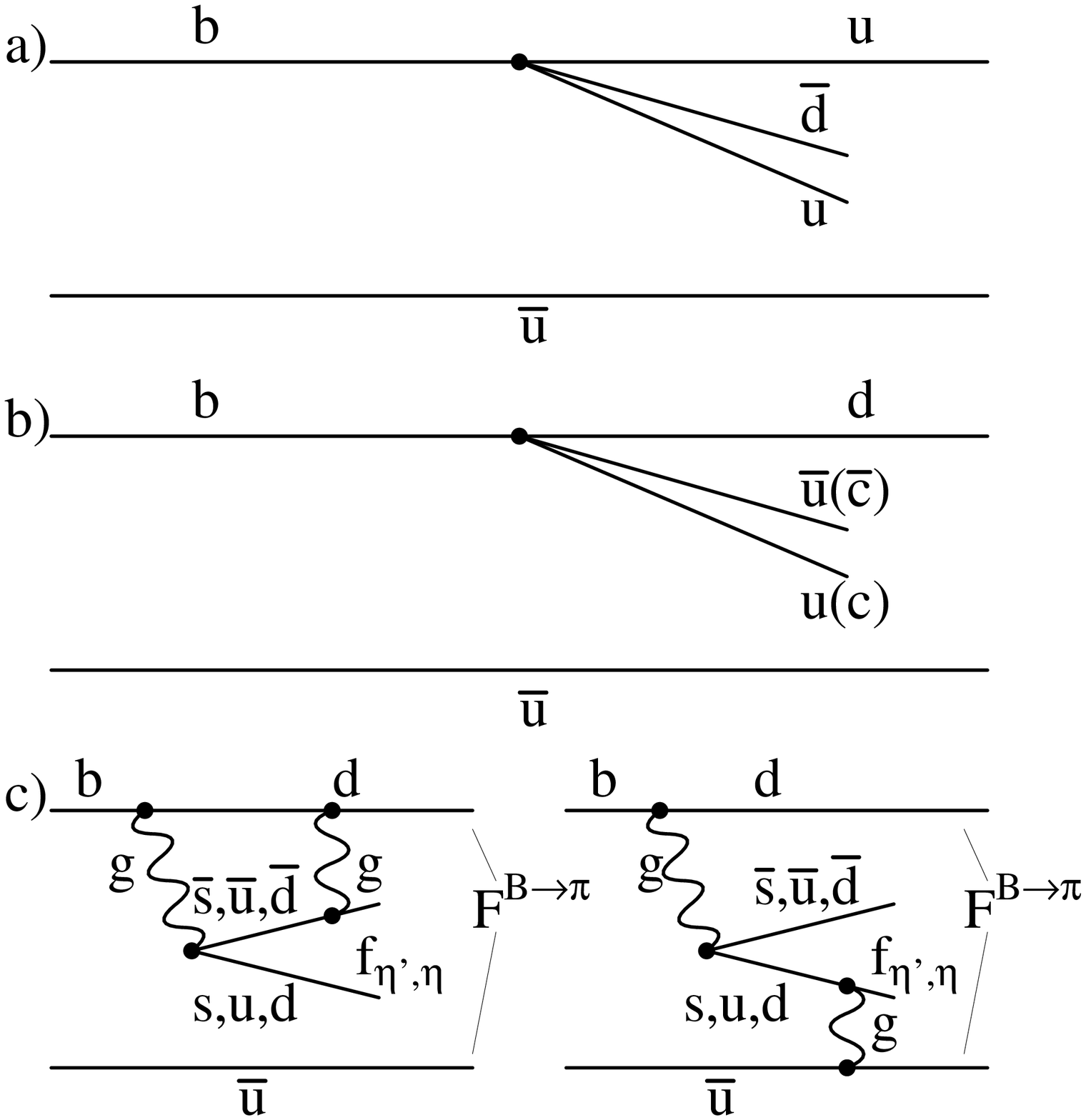}}
\caption{\small a) Quark diagram for $B^{\mp}\to \pi^{\mp}\eta'(\eta)$ from the model for exclusive
matrix elements \cite{ref18,ref19} used in the calculations of reference 7, proportional to $a_1$ in
Eq.~(9).\newline
b) Quark diagram for $B^\mp \to \pi^{\mp}\eta'(\eta)$, proportional to  $a_2$ in Eq.~(9). \newline
c) Quark diagrams for a new enhanced amplitude contributing to  $B^\mp \to \pi^{\mp}\eta'(\eta)$.}
\end{center}
\end{figure}


\begin{thebibliography}{9}
\bibitem{ref1} G.~Sciolla, HFAG, talk at Moriond 2005 
\bibitem{ref2} J.~Ocariz, talk at Moriond 2006; \\http://moriond.in2p3.fr/EW/2006/Transparencies/J.Ocariz.pdf
\bibitem{ref3} Belle Collaboration, Phys.~Rev.~Lett.~{\bf 87} (2001) 091802
\bibitem{ref4} BaBar Collaboration, Phys.~Rev.~Lett.~{\bf 87} (2001) 091801
\bibitem{ref5} J.~Boyd, HFAG, talk at Moriond 2005
\bibitem{ref6} http://www.slac.stanford.edu/xorg/hfag/rare/winter06/acp/index.html
\bibitem{ref7} S.~Barshay, D.~Rein and L.~M.~Sehgal, Phys.~Lett.~{\bf B259} (1991) 475
\bibitem{ref8} S.~Barshay and G.~Kreyerhoff, Mod.~Phys.~Lett.~{\bf A18} (2003) 2887, Appendix; Erratum,
Mod.~Phys.~Lett.~{\bf A19} (2003) 1561
\bibitem{ref9} Review of Particle Properties, Phys.~Lett.~{\bf B592} (2004) 130
\bibitem{ref10} Review of Particle Properties, Phys.~Lett.~{\bf B592} (2004) 156
\bibitem{ref11} M.~Kobayashi and T.~Maskawa, Prog.~Theor.~Phys.~{\bf 49} (1973) 652
\bibitem{ref12} S.~Barshay and G.~Kreyerhoff, Europhysics Lett.~{\bf 63} (2003) 519
\bibitem{ref13} Review of Particle Properties, Phys.~Lett.~{\bf B592} (2004) 296
\bibitem{ref14} T.~Inami and C.~S.~Lim, Prog.~Theor.~Phys.~{\bf 65} (1981) 297, Erratum 1772
\bibitem{ref15} W.~S.~Hou, A.~Soni and H.~Steger, Phys.~Rev.~Lett.~{\bf 59} (1987) 1521
\bibitem{ref16} O.~Nachtmann and C.~Schwanenberger, hep-ph/0308198
\bibitem{ref17} S.~Barshay and G.~Kreyerhoff, Phys.~Lett.~{\bf B578} (2004) 330\\
In line 10 on page 332, correct a misprint to read $A_\eta$.
\bibitem{ref18} M.~Bauer, B.~Stech and M.~Wirbel, Z.~Phys.~{\bf C34} (1987) 103 
\bibitem{ref19} M.~Neubert, B.~Stech, in {\it Heavy Flavours}, A.~J.~Buras, M.~Lindner (Eds.),
World Scientific, Singapore, second ed.; hep-ph/9705292
\bibitem{ref20} Review of Particle Properties, Phys.~Lett.~{\bf B592} (2004) 495
\bibitem{ref21} BaBar Collaboration, B.~Aubert et al., hep-ex/0303039\\
BaBar Collaboration, B.~Aubert et al., hep-ph/0308015
\bibitem{ref22} H.~J.~Lipkin, Phys.~Lett.~{\bf B254} (1991) 247
\end{thebibliography}
\end{document}